\documentclass{article}

\usepackage{arxiv}

\usepackage{bm}
\usepackage{amssymb, amsmath} 
\usepackage{multirow} 

\usepackage{hyperref}       
\usepackage{url}            
\usepackage{booktabs}       
\usepackage{amsfonts}       
\usepackage{nicefrac}       
\usepackage{microtype}      
\usepackage{cleveref}       
\usepackage{lipsum}         
\usepackage{graphicx}
\usepackage{natbib}
\usepackage{doi}

\title{Machine-learning based flow field estimation using floating sensor locations}

\newif\ifuniqueAffiliation
\uniqueAffiliationtrue

\ifuniqueAffiliation 
\author{
	\textbf{Tomoya Oura}\\
	Department of Mechanical Engineering\\
	Keio University\\
	Yokohama, 223-8522, Japan\\
	\texttt{oura.tomoya@keio.jp}\\
	\and
	\textbf{Reno Miura}\\
	Department of Mechanical Engineering\\
	Keio University\\
	Yokohama, 223-8522, Japan\\
	\and
	\textbf{Koji Fukagata}\\
	Department of Mechanical Engineering\\
	Keio University\\
	Yokohama, 223-8522, Japan
}

\renewcommand{\shorttitle}{}

\hypersetup{
pdftitle={A template for the arxiv style},
pdfsubject={q-bio.NC, q-bio.QM},
pdfauthor={David S.~Hippocampus, Elias D.~Striatum},
pdfkeywords={First keyword, Second keyword, More},
}

\begin{document}
\maketitle

\begin{abstract}
	Based on machine learning techniques, we propose a novel method to estimate flow fields using only floating sensor locations. This method does not require either ground-truth velocity fields or governing equations for fluid flows, which is attractive for practical applications. The machine learning model is supposed to generate accurate velocity fields so that the time variation of sensor motion is consistent with the given data of sensor locations. To validate the method, the estimation accuracy, the dependence on the number of sensors, the time intervals for the sensor location data, and the robustness to noise are investigated using three examples of two-dimensional flows: the flow around a circular cylinder, the forced homogeneous isotropic turbulence, and the ocean currents. These investigations demonstrate the performance and practicality of this method, revealing that the accuracy can be comparable to the state-of-the-art physics-informed neural networks (PINNs)-based method even without any assumption of governing equations. Moreover, we observe that the present method can estimate the major structures, such as periodic wakes behind a cylinder, coherent structures in the forced turbulence, and stable ocean currents, with only a few sensors. We believe the present method can provide effective utilization of floating sensor observations in various fields.
\end{abstract}


\section{Introduction}
\label{sec:introduction}

Estimation of turbulent flow fields is an essential issue, particularly in the context of environmental problems these days. For instance, accurate observations of near-surface ocean currents are important for understanding the climate of the earth. To measure flow fields, several projects using floating buoys, called drifters, are globally conducted in oceanography~\cite{wong_argo_2020,hansen_quality_1996}. Because fluid dynamics are nonlinear phenomena, conventional linear methods, such as the linear interpolation and the proper orthogonal decomposition~\cite{mokhasi_predictive_2009,podvin_combining_2018}, have potential drawbacks in the accuracy of the estimation. Thus, effective methods to utilize the sparse sensor measurements are desired.

One such method is data assimilation, which is the process of integrating measurement data into numerical simulations. In data assimilation, model equations cover the nonlinearity, whereas the observations assist the accuracy of model predictions. Using this method, the global ocean state has been estimated with an ocean general circulation model as the nonlinear model equations~\cite{wunsch_practical_2007}. Other studies are also conducted for the river state estimation with two-dimensional shallow water equations and floating sensor measurements~\cite{tossavainen_ensemble_2008,tinka_floating_2013}. The limitation of this method is that the estimation requires an assumption of governing equations for fluid flows.

Another candidate is using machine learning (ML) techniques. Because ML contains nonlinearity as activation functions, it has the ability to handle the nonlinear fluid flow phenomena. Using ML, several studies~\cite{callaham_robust_2019,erichson_shallow_2020} have been conducted to estimate two-dimensional flow fields from sparse sensor measurements for various targets, such as the flow around a circular cylinder, mixing layer, sea surface temperature, forced isotropic turbulence, etc. While the situations in these studies assumed that the sensor locations are fixed through all observations, the method proposed by \cite{fukami_global_2021} can consider unfixed sensor measurements for two- and three-dimensional field estimation using Voronoi tessellation and the convolutional neural networks (CNN)~\cite{y_lecun_gradient-based_1998}. The restriction of all studies above is that the training of ML models is conducted in a supervised manner; in other words, high-resolution flow fields (i.e., ground-truth fields) must be prepared.
\cite{clark_di_leoni_reconstructing_2023}, in contrast, proposed a method to estimate three-dimensional flow fields using physics-informed neural networks (PINNs)~\cite{raissi_physics_2019}. In their method, the networks enforce both the observed sensor velocities and the physical constraints to reconstruct the flow fields under the assumption that the flow is governed by the continuity equation and the Navier--Stokes equations. In addition to the research above, ML has recently gained attention in the field of particle image velocimetry (PIV) as well. Both supervised methods employing CNN with synthesized particle images~\cite{cai_particle_2020,morimoto_experimental_2021,zhang_pyramidal_2023} and the unsupervised methods including image back-warping strategy inspired by computer vision~\cite{Hasanuzzaman_enhancement_2023} or physics-informed loss~\cite{zhang_unsupervised_2023} are proposed for the PIV process. Although PIV does not rely on individual particle positions but utilizes particle images within interrogation windows, these efforts using ML techniques provide valuable insights for the present study.

In this paper, we propose a novel method for estimating flow fields from floating sensor measurements using ML techniques. Our method is based only on sequential data of floating sensor locations, meaning that it can be used without knowing either ground-truth velocity fields or governing equations for fluid flows.
In the present method, the ML model is trained so that the model generates the estimated velocity fields which are consistent with the time variation of the given sensor locations. The present method relies solely on a simple assumption, that is, the equation of motion for floating sensors. The distinctive features of our method are expected to lead us to a more practical use of floating sensor observations in various situations.

The paper is organized as follows. We introduce the concept and structure of our ML model in \S~\ref{sec:concept_and_model_struture}, and three examples of flow fields, including the two-dimensional flow around a circular cylinder, the two-dimensional forced homogeneous isotropic turbulence (HIT), and the two-dimensional ocean currents, for validation in \S~\ref{sec:datasets}. In \S~\ref{sec:result_cylinder} and \S~\ref{sec:result_forced_hit}, some fundamental investigations are conducted using the two-dimensional flow around a cylinder and the two-dimensional forced HIT. Subsequently, we demonstrate the practical applicability of our method using the dataset of the two-dimensional ocean currents in \S~\ref{sec:result_ocean_currents}. Finally, concluding remarks are presented in \S~\ref{sec:conclusions}.

\section{Methods}
\label{sec:methods}

\subsection{Concept and model structure}
\label{sec:concept_and_model_struture}

\begin{figure}
  \centerline{
    \includegraphics[width=120mm]{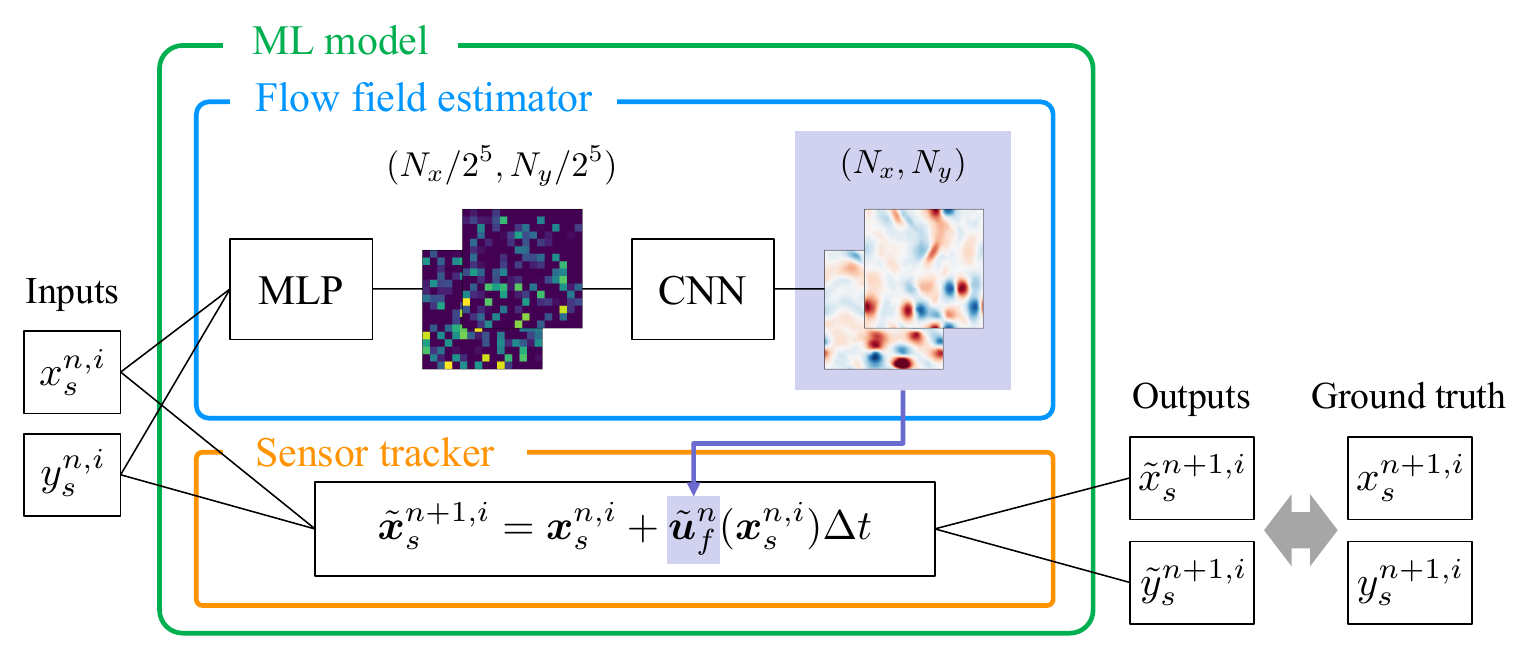}
  }
  \caption{The schematic drawing of the present method.}
  \label{fig:model_schematics}
\end{figure}

The present method, which is based on the ML techniques, contains two parts in the model: the flow field estimator and the sensor tracker. Figure~\ref{fig:model_schematics} shows the schematic of the present method. The required data for training of the model is only sequential sensor locations $\{ \boldsymbol{x}_s^{n, i} \} \in \mathbb{R}^{N_t \times N_s \times d}$, with a fixed time-step size $\Delta t$. Here $N_t$ is the number of time steps, $N_s$ is the number of sensors, and $d$ is the dimension of flow fields. In the case of two-dimensional flow ($d = 2$), $\boldsymbol{x}_s^{n, i}$ can be represented as $(x_s^{n, i}, y_s^{n, i})$. The superscript $n$ means the time step of the sequential data and $i$ is the sensor indices. These superscripts will be omitted in some contexts for a notational simplicity.

When receiving the sensor locations, each part of the model acts as follows. First, the flow field estimator converts the given $\boldsymbol{x}_s^{n, i}$ to the velocity-field-like arrays through the operation of the multi-layer perceptron (MLP)~\cite{rumelhart1986learning} and several filters and upsampling operators of the convolutional neural networks (CNN). Subsequently, the sensor tracker supposes the arrays obtained by the flow field estimator to be the flow velocity fields at $t = n \Delta t$, and conducts time integration of sensor locations. As a result, the ML model outputs the estimated sensor locations at $t = (n + 1) \Delta t$, denoted as $\tilde{\boldsymbol{x}}_s^{n+1, i}$.

In the phase of model training, the estimated sensor locations $\tilde{\boldsymbol{x}}_s^{n+1, i}$, are compared to the ground truth of the sensor locations $\boldsymbol{x}_s^{n+1, i}$ at $t = (n + 1) \Delta t$. The objective of the training is to optimize the trainable weights in the flow field estimator so that the discrepancy between $\tilde{\boldsymbol{x}}_s^{n+1, i}$ and $\boldsymbol{x}_s^{n+1, i}$ becomes minimum, namely,
\begin{equation}
  \boldsymbol{w} = {\textrm{argmin}}_{\boldsymbol{w}} \left\| \boldsymbol{x}_s^{n+1, i} - \tilde{\boldsymbol{x}}_s^{n+1, i} \right\|_2^2
  ,
  \label{eq:weight_optimisation}
\end{equation}
where $\boldsymbol{w}$ represents the trainable weights of MLP and CNN in the flow field estimator. After all weights $\boldsymbol{w}$ are optimized, the flow field estimator is expected to output the accurate flow velocity fields at $t = n \Delta t$.

The detail of the model structure is explained as follows. Hereafter, only the case of the two-dimensional flow (i.e., $d = 2$) is considered. In the flow field estimator, MLP contains four hidden layers with 256 nodes each. MLP obtains the vector of sensor locations $\boldsymbol{x}_s^{n, i} \in \mathbb{R}^{N_s \times 2}$, and outputs the array whose size is $(N_x / 2^5, N_y / 2^5, 2)$, where $(N_x, N_y)$ is the number of grid points of target velocity fields. This array is then passed to the CNN part. The CNN part contains five sets of convolution layers, which have 48 filters with the kernel size of $(7, 7)$, and upsampling layers, whose size is $(2, 2)$. As the output layer, there is an additional convolutional layer at the bottom of CNN. The Rectified Linear Unit (ReLU) is adopted for the activation function in both MLP and CNN except for the output layer where the linear activation is employed. In the convolutions, the periodic padding and the replication padding are used for the periodic and the other flow directions, respectively. As a result, the flow field estimator outputs a two-dimensional array whose size is $(N_x, N_y)$ with two channels corresponding to two velocity components. In the sensor tracker, sensor locations at the next time step are calculated using the estimated fluid velocity field. The numerical scheme for the sensor tracking will be explained in \S~\ref{sec:sensor_tracking_method}.

For the model training, Adam~\cite{kingma2017adam} is adopted as the weight optimizer, and the loss is evaluated with the mean squared error. The number of epochs is set to $2000$ based on the loss convergence, and the model with the minimum loss during the training is used for validation. All training in this study is conducted with an NVIDIA Tesla A100 graphics processing unit. It should be noted here that the purpose of the present method is to \textit{estimate} the flow fields from the \textit{observed} sensor locations. Unlike typical ML applications, the model does not need to \textit{predict} flow fields with \textit{unseen} sensor locations. Thus, the whole dataset for all time periods can be supplied simultaneously for model training and there is no need to separate the dataset for training and testing.

\subsection{Datasets}
\label{sec:datasets}

For the validation of the present method, here three kinds of flow are prepared: the two-dimensional flow around a circular cylinder, the two-dimensional forced HIT, and the two-dimensional ocean currents. Some details on these fields are explained in this subsection.

\subsubsection{Two-dimensional flow around a circular cylinder}
\label{sec:method_cylinder}

A low-Reynolds number two-dimensional flow around a circular cylinder is used to introduce how the present method performs. The governing equations are the incompressible continuity and Navier--Stokes equations:
\begin{eqnarray}
  \nabla \cdot \boldsymbol{u}_f &=& 0
  ,\\
  \label{eq:continuity_eq}
  \frac{\partial \boldsymbol{u}_f}{\partial t} &=& - \nabla \cdot (\boldsymbol{u}_f \boldsymbol{u}_f) - \nabla p_f + \frac{1}{Re} \nabla^2 \boldsymbol{u}_f
  ,
  \label{eq:NS_eq}
\end{eqnarray}
where $\boldsymbol{u}_f$ is the fluid velocity, $p_f$ is the pressure. Here all variables are nondimensionalized with the uniform inflow velocity and the cylinder diameter, and the Reynolds number $Re$ is defined using these and the kinematic viscosity. The direct numerical simulation (DNS) is conducted using the second-order central finite difference method and the ghost-cell immersed boundary method at $Re = 100$. The size of the computational domain is set to $25.6$ and $20.0$ for the streamwise ($x$) and transverse ($y$) directions, respectively. The center of the cylinder, hereafter defined as the origin of the coordinates, is located at the distance of $9$ away from the inflow boundary. For more details on the DNS, refer to \cite{kor_unified_2017}. The region of interest (ROI) of field estimation is set to $[-1.0, 5.4]$ and $[-1.6, 1.6]$ for $x$ and $y$ directions, respectively, and the time range is set to $t = 0$\,--\,$31.25$ with the time step size of $\Delta t_f = 2.5 \times 10^{-3}$. This time range is about five times longer than the vortex shedding period. The number of grid points in ROI is $N_x = 256$ and $N_y = 128$.

\subsubsection{Two-dimensional forced homogeneous isotropic turbulence (HIT)}
\label{sec:method_forced_hit}

The two-dimensional forced HIT, which is stationary turbulence, is used for the basic investigation of the present method. In nature, fluid flow whose depth is much less than horizontal scale, such as atmosphere and ocean, can be regarded as two-dimensional~\cite{charney_geostrophic_1971,xia_spectrally_2009,san_stationary_2013}.

The governing equation is the two-dimensional vorticity equation derived from the incompressible continuity and Navier--Stokes equations. In numerical simulations of two-dimensional stationary turbulence, two artificial terms, which are the forcing term and damping term, are required to prevent the total energy dissipation and the inverse energy cascade~\cite{san_stationary_2013}. The equation is represented as
\begin{equation}
  \frac{\textrm{D} \omega_{f}}{\textrm{D}t} = \nu \nabla^2 \omega_{f} + F + D
  ,
  \label{eq:vorticity_eq}
\end{equation}
where $\omega_{f}$ is the vorticity, $\textrm{D}/\textrm{D}t$ is the material derivative, $\nu$ is the kinematic viscosity, $F$ is the forcing term, and $D$ is the damping term. For the forcing term, inspired by ~\cite{lundgren_linearly_2003}, the band-limited linear forcing is used. The forcing term is proportional to the vorticity itself and acted within the limited wave number $k_{F,\textrm{min}} < k < k_{F,\textrm{max}}$. Similarly, for the damping term, the band-limited linear damping, which is adopted in many studies~\cite{maltrud_energy_1991,schorghofer_energy_2000,san_stationary_2013}, is used. The damping term is also linear to the vorticity within the limited wave number, but the scale factor $\gamma$ and the band range $k_{D,\textrm{min}} < k < k_{D,\textrm{max}}$ are generally different from ones of the forcing term. In sum, the forcing and damping terms are represented as 
\begin{equation}
  F = \left\{
    \begin{array}{ll}
      A \omega_{f}, & k_{F,\textrm{min}} < k < k_{F,\textrm{max}} \\[2pt]
      0,         & \textrm{otherwise,}
    \end{array} \right.
  \label{eq:forcing_term}
\end{equation}
and
\begin{equation}
  D = \left\{
    \begin{array}{ll}
      - \gamma \omega_{f}, & k_{D,\textrm{min}} < k < k_{D,\textrm{max}} \\[2pt]
      0,         & \textrm{otherwise,}
    \end{array} \right.
  \label{eq:damping_term}
\end{equation}
respectively. In this study, the kinematic viscosity is set to $\nu = 5 \times 10^{-4}$, the forcing term is acted with $A = 0.075$ in the range of $4 \leq k < 6$, and the damping term is acted with $\gamma = 0.1$ in the range of $0 < k \leq 3$. For the numerical setup, the domain size is set to $2 \pi$ for both $x$ and $y$ directions with $N_x = N_y = 512$ grid points in each, and the time step $\Delta t_f$ is set to $10^{-3}$. The initial field is generated with randomly distributed 100 Taylor vortices. The boundary conditions are periodical in both directions. The Taylor Reynolds number, based on the root mean square (RMS) of velocity fluctuations $u_{\textrm{rms}}$, the Taylor microscale $\lambda_{g}$, and the kinematic viscosity, becomes $Re_{\lambda} \simeq 272$ in the stationary state. The direct numerical simulation is conducted with the Fourier spectral method and the fourth-order Runge--Kutta method following \cite{taira_network_2016}.

\subsubsection{Two-dimensional ocean currents}
\label{sec:method_ocean_currents}

To demonstrate the practicality of the present method, two-dimensional ocean current fields are considered for a practical target. The dataset is obtained with the ocean general circulation model
for the earth simulator (OFES), which is the eddy-resolving ocean simulations conducted on the Earth Simulator under the support of JAMSTEC~\cite{masumoto_fifty_2004,sasaki_series_2004,sasaki_chloro_2004,sasaki_eddy_2008}. OFES provides the daily mean velocity fields with the grid sizes of $0.1 ^\circ \textrm{N}$ and $0.1 ^\circ \textrm{E}$ in the world ocean.

In this study, the two-dimensional velocity fields (i.e., the zonal velocity $u_f$ and the meridional velocity $v_f$) at the water depth of $2.5$ m in the eastern ocean of Japan are used. ROI is set to $137.3$\,--\,$162.8 ^\circ \textrm{E}$ and $33.6$\,--\,$39.9 ^\circ \textrm{N}$, where the number of grid points is $N_x = 256$ for longitude and $N_y = 64$ for latitude. The target period is from 1 January 1991 to 31 December 1991. In the original dataset, the time interval of each field is 24 hours because all data is daily averaged. For appropriate sensor tracking, the sixth-order Lagrangian interpolation is employed in time direction, resulting in the time step of $\Delta t_f = 1$ hour. 

\subsection{Sensor tracking method}
\label{sec:sensor_tracking_method}

For numerical simulations of sensor locations in flow fields, Lagrangian particle tracking, which is the well-known method for tracking discrete particles in a continuous phase, is adopted. In the present study, the sensors are regarded as tracer particles; namely, the particle velocity is equal to the fluid velocity at the particle locations, and the particle number density is low enough to ignore the particle-to-fluid and particle-to-particle interactions. This assumption can be justified when the particle Stokes number is sufficiently smaller than unity.
Considering a floating buoy with a size of $30$~cm in water, for example, the particle Stokes number becomes smaller than unity when the characteristic flow length is roughly larger than $\sim 10$~km, as in the case of wide ocean currents.
The equation of motion for sensors is represented as
\begin{equation}
  \frac{d \boldsymbol{x}_s}{dt} = \boldsymbol{u}_f (\boldsymbol{x}_s)
  .
  \label{eq:equation_of_motion_for_sensors}
\end{equation}
In this study, the fluid velocity at a sensor location $\boldsymbol{u}_f (\boldsymbol{x}_s)$ is calculated with the bilinear interpolation from the estimated velocity field, and the time integration of equation (\ref{eq:equation_of_motion_for_sensors}) is performed with the Euler explicit method for an implementation simplicity. The time step size of sensor tracking $\Delta t$ is set to $\Delta t = 5 \Delta t_f$ for the flow around a cylinder (\S~\ref{sec:method_cylinder}), $\Delta t = 10 \Delta t_f$ for the forced HIT (\S~\ref{sec:method_forced_hit}), and $\Delta t = \Delta t_f$ for the ocean currents (\S~\ref{sec:method_ocean_currents}). Note that, in the case of the forced HIT, we have confirmed no differences between the results of $\Delta t = \Delta t_f$ and $\Delta t = 10 \Delta t_f$. In terms of boundaries, the periodic condition is applied in the case of the forced HIT, whereas the sensors that reach outside the ROI or inside the non-fluid area (i.e., the cylinder body and the islands) are relocated in the case of the flow around a cylinder and the ocean currents. The relocation points are selected at $x = -1.0$ and random $y$ for the flow around a cylinder, and at random $x$ and $y$ for the ocean currents.

\section{Results and discussion}
\label{sec:results_and_discussion}

\subsection{Example 1: Two-dimensional flow around a circular cylinder}
\label{sec:result_cylinder}
\begin{figure}
  \centerline{
    \includegraphics[width=120mm]{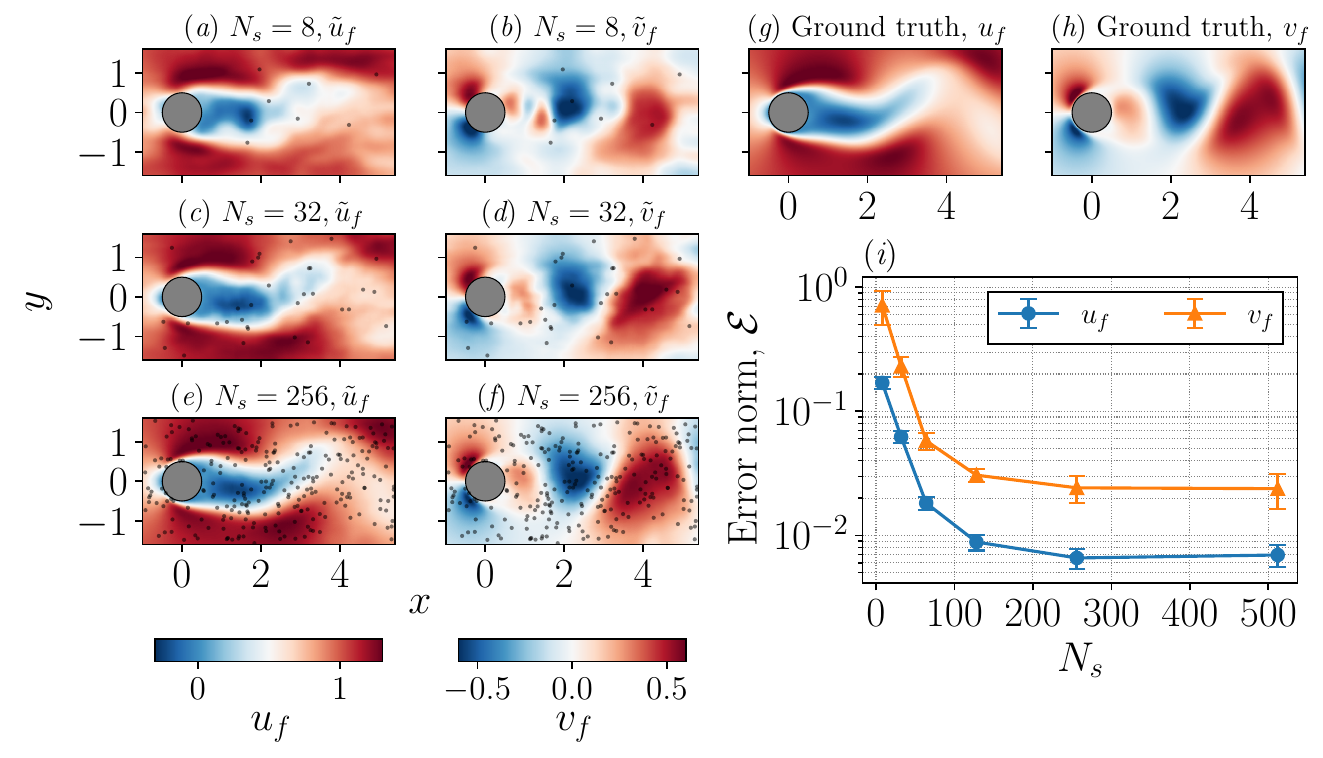}
  }
  \caption{The dependence on the number of sensors $N_s$ in the case of the flow around a circular cylinder. (\textit{a}--\textit{h}) The estimated and ground-truth velocity fields. The grey dots in (\textit{a}--\textit{f}) represent the sensor locations. (\textit{i}) The error norm.}
  \label{fig:cylinder_n_sensors_results}
\end{figure}
First, we quickly demonstrate the estimation performance of the present method using the flow around a circular cylinder. Hereafter, we will compare the estimated flow fields with the ground truth fields. It should be emphasized again that the ground truth fields are prepared only to validate the estimation performance and are not used to train the ML model. This assumes a practical situation where the sensor locations can be observed, for example, using a global positioning system (GPS), whereas the velocity fields between sensors are unknown and need to be estimated.

Figure~\ref{fig:cylinder_n_sensors_results} shows the ground truth and the estimated flow fields for different numbers of sensors $N_s$. Compared with the ground truth fields, the velocity fields are almost perfectly reconstructed with $N_s = 256$. This result suggests the significant potential of the present method. Moreover, the major flow structures can be estimated with a small number of sensors, even in the regions without sensors. This is possibly due to the following reasons. The wake of a cylinder has temporal periodicity. The ML model is simultaneously trained with the sequential data and the weights in the flow field estimator are optimized over the entire dataset. Thus the ML model can estimate the flow fields with even a particularly small number of sensors such as $N_s = 8$.

For the quantitative evaluation, the normalized error norm is defined as
\begin{equation}
  \mathcal{E} = \frac{\left\| u_f - \tilde{u}_f \right\|_F}{\left\| u_f \right\|_F}
  ,
  \label{eq:normalized_error_norm}
\end{equation}
where $\left\| \cdot \right\|_F$ is the Frobenius norm. In this study, the error norm is calculated for each instantaneous field and the mean and standard deviation of them will be presented. Figure~\ref{fig:cylinder_n_sensors_results}~(\textit{i}) shows $N_s = 256$ is sufficient to estimate the flow fields in this case. The mean sensor distances $\delta$ normalized with the cylinder diameter is $0.28$ in the case of $N_s = 256$.

\subsection{Example 2: Two-dimensional forced HIT}
\label{sec:result_forced_hit}

More comprehensive investigation of the present method is performed using the two-dimensional forced HIT. First, the verification of the dataset is conducted. Subsequently, the dependence on the number of sensors, the time intervals of given data, and the robustness to noise are considered. Finally, we compare the proposed method with existing methods and demonstrate the advantages of the present method.

\subsubsection{Verification of the flow fields}
\label{sec:forced_hit_verification}
\begin{figure}
  \centerline{
    \includegraphics[width=120mm]{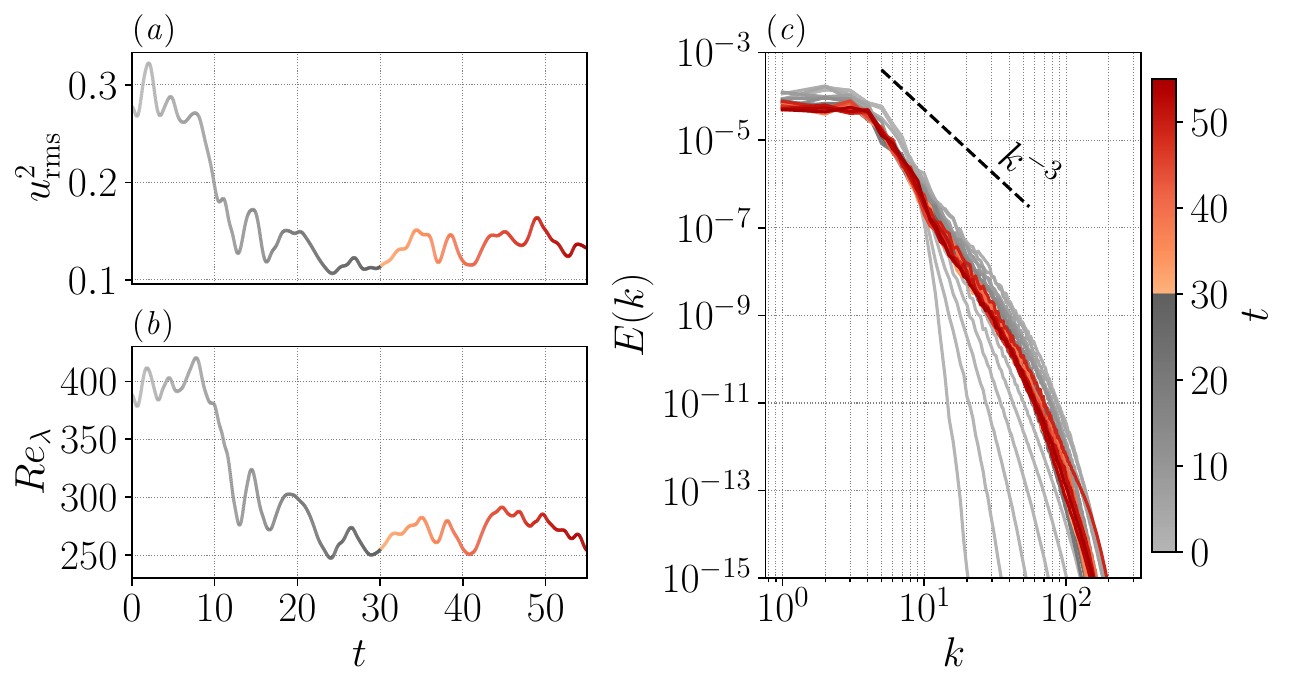}
  }
  \caption{The statistics of the forced HIT: (\textit{a}) $u_{\textrm{rms}}^2$, (\textit{b}) the Taylor Reynolds number, and (\textit{c}) the energy spectrum. The warm-colored range is adopted for the validation.}
  \label{fig:forced_hit_stats}
\end{figure}
To verify the flow fields, the statistics, including $u_{\textrm{rms}}^2$, the Taylor Reynolds number, and the energy spectrum of the instantaneous fields, are presented in figure~\ref{fig:forced_hit_stats}. From the results, all these statistics are converged after $t \sim 30$. One may realise that, in the energy spectrum for $t > 30$, the slope in the inertial range is steeper than $k^{-3}$. Although the $k^{-3}$ law was theoretically argued with dimensional analysis~\cite{leith_diffusion_1968}, several literature~\cite{legras_high-resolution_1988,maltrud_energy_1991,san_stationary_2013} reported that the slope can be steeper than $k^{-3}$ depending on the forcing parameters. Thus, the time range for field estimation is determined to be $30 \leq t < 55$, shown as warm-colored lines in figure~\ref{fig:forced_hit_stats}.

\subsubsection{The dependence on the number of sensors}
\label{sec:forced_hit_number_of_sensors}
\begin{figure}
  \centerline{
    \includegraphics[width=135mm]{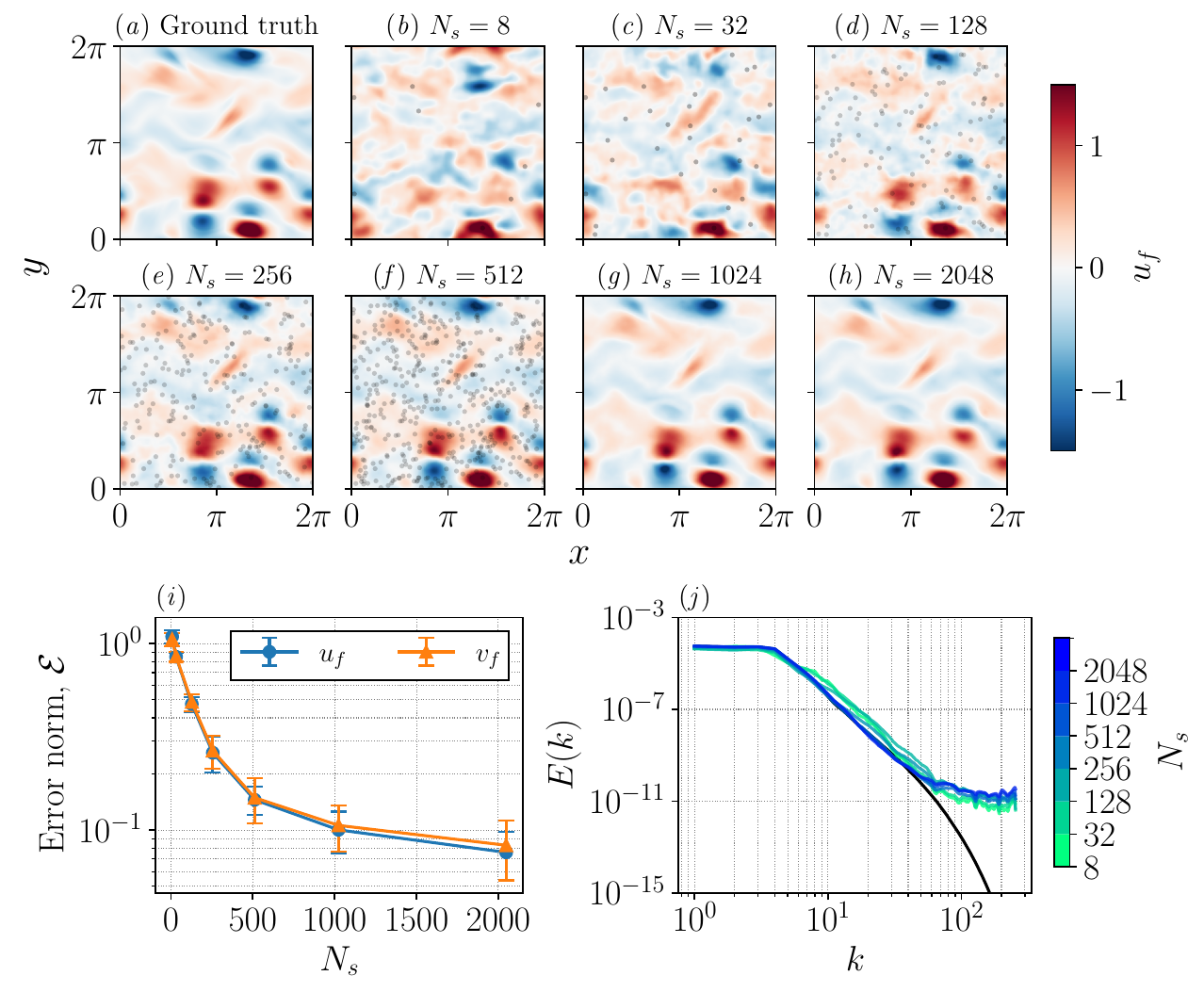}
  }
  \caption{The dependence on the number of sensors $N_s$ in the case of the forced HIT. (\textit{a}--\textit{h}) The ground-truth and estimated velocity fields at $t = 40.24$. The grey dots in (\textit{b}--\textit{f}) represent the sensor locations. In (\textit{g}) and (\textit{h}), the sensor locations are not printed for better readability. (\textit{i}) The error norm. (\textit{j}) The energy spectrum, where the black solid line shows the ground truth.}
  \label{fig:forced_hit_n_sensors}
\end{figure}

To investigate the influence of the number of sensors $N_s$, the training and evaluation of the ML model using different $N_s$ are conducted. Figure~\ref{fig:forced_hit_n_sensors} shows the instantaneous velocity fields estimated using different $N_s$. Note that hereafter only the $x$-component of the velocity is shown for brevity, but the same trend is observed for the $y$-component. The dot plots in figure~\ref{fig:forced_hit_n_sensors} (\textit{b}--\textit{f}) represent the sensor locations at the time step, whereas these are not printed in (\textit{g}) and (\textit{h}) for better readability. Figure~\ref{fig:forced_hit_n_sensors} (\textit{i}) and (\textit{j}) also present the time-averaged error norm and the energy spectrum depending on the number of sensors for a quantitative analysis. These results suggest that 512 or more sensors can estimate the velocity fields and the energy spectrum in $k \lesssim 40$ almost perfectly in this problem setting, although the smaller structures are lost.

Moreover, the remarkable attribute is observed with a small number of sensors. Focusing on the region of the large velocity magnitude (e.g., the dark-colored region around $(x, y) = (4 \pi / 3, \pi / 10)$ in figure~\ref{fig:forced_hit_n_sensors} (\textit{a}--\textit{h})), the flow structures can be reasonably well reconstructed even when there are only a few sensors around the region. In two-dimensional turbulence, strong eddies called coherent structures, represented as the doublets of positive and negative velocities, appear with relatively long lifetimes. If there is no sensor around the eddies at a certain time step, sensors can move around the region in other time steps. CNN is generally recognized for the ability to interpolate unobserved regions. In image processing, CNN-based models have demonstrated effectiveness in completing missing regions of unseen pictures through the context of the training data~\cite{pathak_context_2016,iizuka_globally_2017}. This capability has been applied to the PIV process by \cite{morimoto_experimental_2021}. Consequently, the ML model may learn the coherent structure through the training process and be able to estimate the velocity fields around the structure without sensors at a certain time step. This feature is attractive in practical applications with statistically steady flow structures, which will be discussed in \S~\ref{sec:result_ocean_currents}.

\subsubsection{The dependence on time intervals}
\label{sec:forced_hit_data_interval}
\begin{figure}
  \centerline{
    \includegraphics[width=120mm]{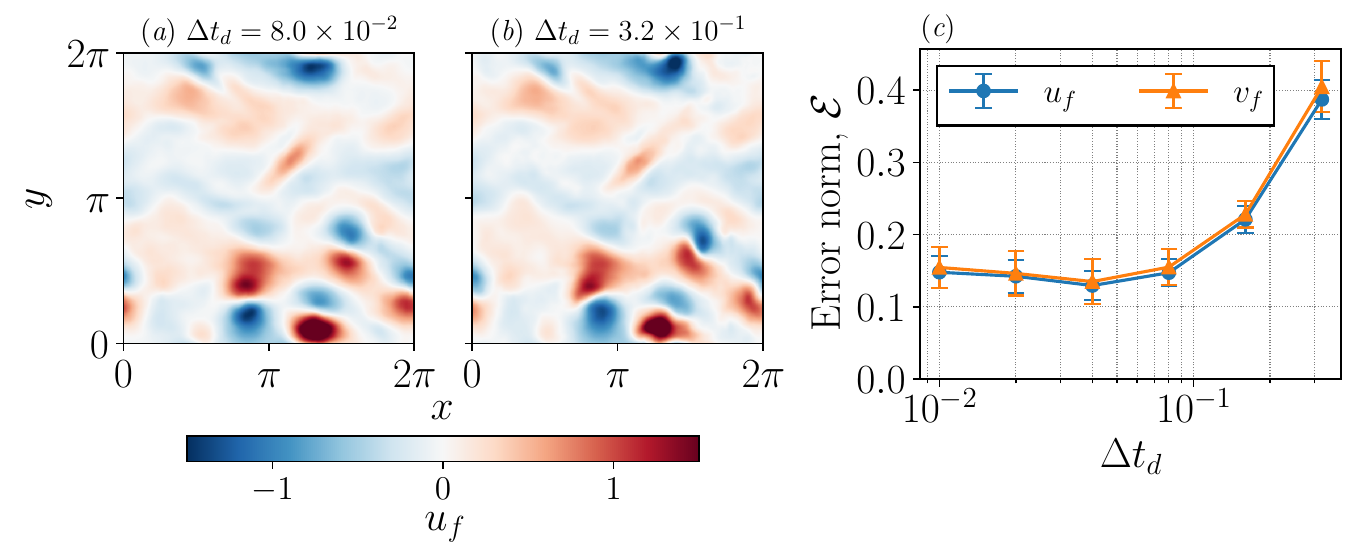}
  }
  \caption{The dependence on the time intervals for the sensor location data in the case of the forced HIT. (\textit{a}--\textit{b}) The estimated velocity fields at $t = 40.24$ using different time intervals~$\Delta t_d$. (\textit{c}) The error norm.}
  \label{fig:forced_hit_data_interval}
\end{figure}
In a practical situation, observations may only be available at large time intervals. Here, the dependence on the time intervals of the sensor location data is investigated, namely, only the data of sensor locations with a certain time interval $\Delta t_d$ are used to train the model. The interval is set to $\Delta t_d = N \Delta t$ for $N = 1, 2, 4, 8, 16, 32$, where $\Delta t$ is the original time step size of the prepared dataset. For a fair comparison, the amount of data used for model training is set to be equal to that in the case of the largest $\Delta t_d$ in this investigation.

Figure~\ref{fig:forced_hit_data_interval} shows the dependence of the estimated velocity fields and the estimation accuracy on the time intervals. The number of sensors is set to $N_s = 512$. Note that $\Delta t_d = 1.0 \times 10^{-2}$ is equal to the original time step size $\Delta t$. In figure~\ref{fig:forced_hit_data_interval}~(\textit{c}), the estimation accuracy can be maintained for $\Delta t_d \lesssim 10^{-1}$, which corresponds to $\Delta t_d u_{\rm rms} / \lambda_g \lesssim 0.1$ in physical meaning. As seen in figure~\ref{fig:forced_hit_data_interval}~(\textit{b}), the velocity field with $\Delta t_d = 3.2 \times 10^{-1}$, which is $32$ times larger than the original time step size $\Delta t$, become slightly distorted but the flow structures can still be estimated in acceptable accuracy. These results suggest the robustness of the present method to the large time intervals of observation data.

\subsubsection{Robustness to noise}
\label{sec:forced_hit_robustness_to_noise}
\begin{figure}
  \centerline{
    \includegraphics[width=130mm]{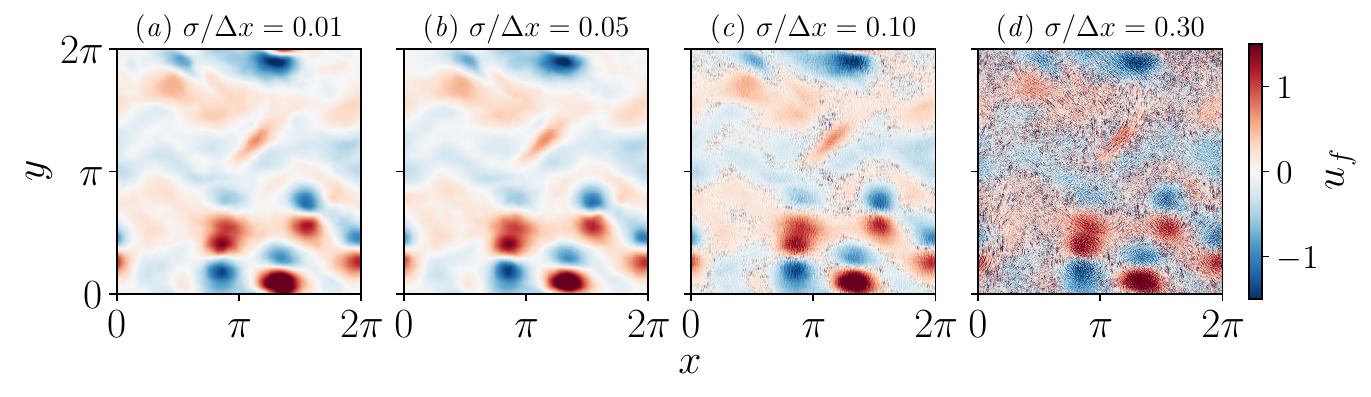}
  }
  \caption{The estimated velocity fields of the forced HIT at $t = 40.24$ depending on the noise intensity $\sigma / \Delta x$.}
  \label{fig:forced_hit_noise_uf_fields}
\end{figure}
Assuming the signal noise of sensors in a practical application, here the Gaussian random noise is applied to the data of sensor locations $\{ \boldsymbol{x}_s^{n, i} \}$. The noise intensity is determined to be $\sigma / \Delta x = 0.01, 0.05, 0.1$ and $0.3$, where $\sigma$ is the standard deviation of the Gaussian distribution and $\Delta x = 2 \pi / N_x$ is the grid size of the flow fields. In physical meaning using the Taylor microscale $\lambda_{g}$, the noise intensities above correspond to $\sigma / \lambda_{g} = 3.3 \times 10^{-4}, 1.7 \times 10^{-3}, 3.3 \times 10^{-3}$ and $9.9 \times 10^{-3}$, respectively. Figure~\ref{fig:forced_hit_noise_uf_fields} shows the instantaneous velocity fields estimated with the noisy data of sensor locations. This result implies that the present ML model can accept noise whose intensity is less than 10\% of the grid size, whereas larger noise deteriorates the accuracy of the estimated velocity fields. This may be due to the fact that the present method relies only on sensor locations. 

\subsubsection{Comparison with existing methods}
\label{sec:forced_hit_existing_methods}
\begin{figure}
  \centerline{
    \includegraphics[width=\linewidth]{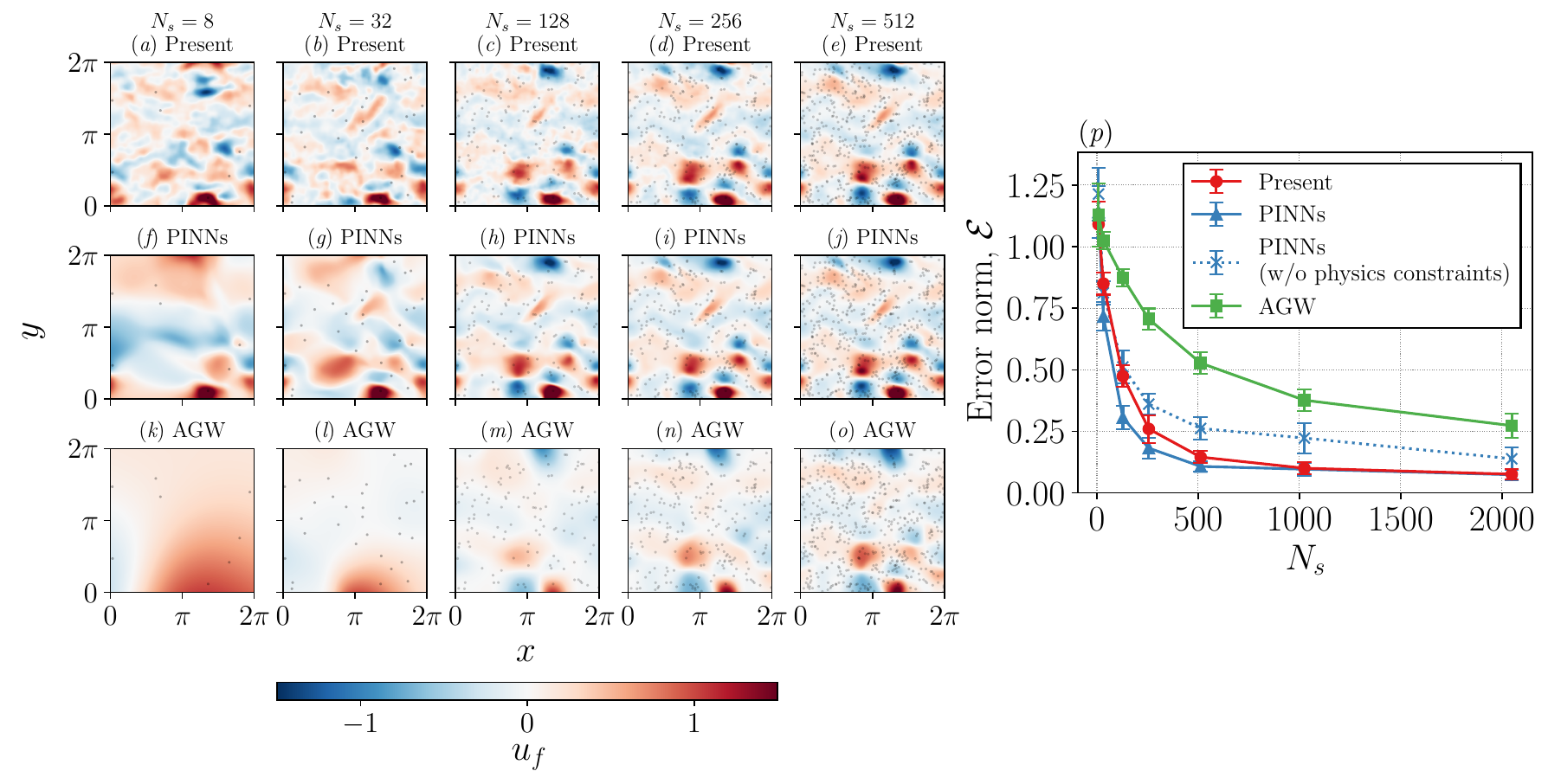}
  }
  \caption{The comparison among the present method, the PINNs-based method, and AGW. (\textit{a}--\textit{o}) The estimated velocity fields of the forced HIT at $t = 40.24$ depending on the number of sensors $N_s$. The grey dots represent the sensor locations. (\textit{p}) The error norm.}
  \label{fig:forced_hit_existing_methods}
\end{figure}
To state the advantages of the present method, we demonstrate the comparison with two existing methods: (1) the adaptive Gaussian windowing (AGW) as a historically established method and (2) the PINNs-based model as a state-of-the-art ML method. 

AGW~\cite{agui_performance_1987}, which is widely used for particle tracking velocimetry (PTV), is a linear interpolation method with Gaussian kernels. Using the observed velocities at each sensor locations $\boldsymbol{u}_s^{n, i} ( \boldsymbol{x}_s^{n, i} )$, the fluid velocity at coordinates $\boldsymbol{x}$ is calculated as
\begin{equation}
  \boldsymbol{u}_f^n (\boldsymbol{x}) = \left. \sum_i \alpha^{n, i} \boldsymbol{u}_s^{n, i} \left(\boldsymbol{x}_s^{n, i} \right) \middle/ \sum_i \alpha^{n, i} \right.
  ,
  \label{eq:AGW_1}
\end{equation}
where $\alpha^{n, i}$ is the Gaussian kernel represented as
\begin{equation}
  \alpha^{n, i} = \exp \left( \frac{- \left| \boldsymbol{x} - \boldsymbol{x}_s^{n, i} \right|^2}{H^2} \right)
  .
  \label{eq:AGW_2}
\end{equation}
The optimum window width~$H$ is determined as $1.24 \delta$ using the mean sensor distances~$\delta$ by \cite{agui_performance_1987}, thus the same parameter~$H$ is adopted here.

As a state-of-the-art ML method, the PINNs-based model proposed by \cite{clark_di_leoni_reconstructing_2023} is selected. In their report, MLP including eight hidden layers with 200 nodes each obtains time and coordinates as the input, and outputs velocities and pressure. 
The networks are optimized so that the output velocities coincide with the given sensor velocities and also follow the governing equations which are assumed as the continuity equation and the three-dimensional Navier--Stokes equations in their study. The more details are explained in \cite{clark_di_leoni_reconstructing_2023}. In the present study, we utilized the code made available by the author \cite{clark_pinns_2022}, while adapting it to the two-dimensional Navier--Stokes equations.

The estimation results with the present method, the PINNs-based method, and AGW depending on the various number of sensors $N_s$ are shown in figure~\ref{fig:forced_hit_existing_methods}.
For reference, the results of the PINNs-based model trained using only sensor observations are also shown in figure~\ref{fig:forced_hit_existing_methods}~(\textit{p}), indicating the importance of physics constraints in the model.
The present method and the PINNs-based method show better accuracy than AGW. Regarding the number of sensors, even $N_s = 2048$ seems insufficient for AGW in figure~\ref{fig:forced_hit_existing_methods}~(\textit{p}), whereas the accuracy of the present and PINNs-based methods is saturated in the case of $N_s = 1024$. The comparison between the present method and the PINNs-based method shows no significant differences when $N_s \geq 256$. This result indicates that the present method achieves an accuracy comparable to the PINNs-based method, despite that the governing equations for fluid flows are not assumed in the model. 

In the remainder of this subsection, three aspects are compared between the present method and the PINNs-based method. First, physics enforcement enables PINNs to estimate both accurate and physically consistent flow fields. However, when the governing equation is completely unknown, this advantage of PINNs becomes limited. Even in such situations, the present method is still applicable. The specific situation will be discussed in \S~\ref{sec:result_ocean_currents}. Moreover, when the governing equation is partially known, the present method can also provide the physically consistent --- specifically, the divergence-free --- estimation. The detail will be presented in \ref{app:divergence_free_constraint}.
Second, the PINNs-based method is flexible in the number of sensors during both training and estimation. In contrast, the present method requires a fixed input size, meaning it cannot handle variations in the number of sensors. Nonetheless, the assumption of a constant number of sensors is reasonable in certain contexts, such as applications involving floating sensors on the ocean surface.
Third, in terms of the computational time for model training, the present method takes 17 hours 46 minutes, while the PINNs-based method takes 4 hours and 29 minutes in the case of $N_s = 512$. The longer computational time for the present method is attributed to the fact that the number of trainable parameters of the present method is approximately three times larger than that of the PINNs-based method in this assessment, although the computational time for the present method is still within a acceptable range.

\subsection{Example 3: Two-dimensional ocean currents}
\label{sec:result_ocean_currents}

\subsubsection{The dependence on the nuber of sensors}
\label{sec:ocean_currents_n_sensors}
\begin{figure}
  \centerline{
    \includegraphics[width=150mm]{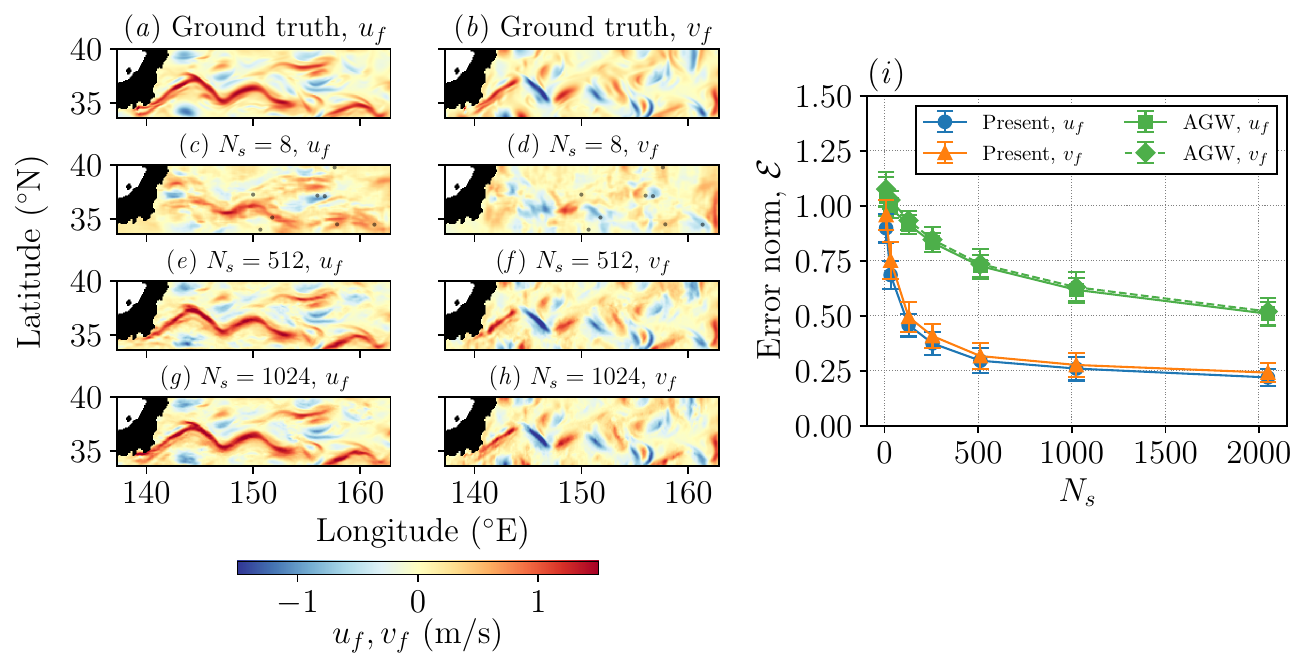}
  }
  \caption{(\textit{a}--\textit{h}) The ground truth and estimated velocity fields of the ocean currents for the number of sensors $N_s = 8$, $512$, and $1024$. The grey dots in (\textit{c}--\textit{d}) represent the sensor locations. In (\textit{e}--\textit{h}), the sensor locations are not printed for better readability. (\textit{i}) The error norm compared with AGW.}
  \label{fig:ocean_n_sensors_results}
\end{figure}
Finally, the investigation using the dataset of ocean currents is conducted to demonstrate the practical applicability. Figure~\ref{fig:ocean_n_sensors_results} shows the results of the estimated velocity fields and the estimation accuracy. Note that the ROI of the estimation is set to the eastern ocean of Japan. The western ocean (i.e., the white-filled region in figure~\ref{fig:ocean_n_sensors_results}~(\textit{a}--\textit{h})) and the islands (i.e., the black-filled region) are omitted when the error norm is calculated. From these results, the velocity fields can be estimated accurately with the 512 sensors, and the error norm is almost saturated with this number of sensors. Note that the computational time for model training is 2 hours 51 minutes in the case of $N_s = 512$.

The comparison with the existing method is also shown in figure~\ref{fig:ocean_n_sensors_results}~(\textit{i}). Here again, the restriction of the PINNs-based method should be commented on. In the case of free-surface flow, the fluid motion on the surface is a consequence of complex three-dimensional motion affected by different factors such as solar heat, Earth's rotation, atmospheric flows, undersea topography, and density variation due to salinity and temperature. Therefore, it is not straightforward to describe the equation of motion on the surface only. As a result, the PINNs-based method described in \S~\ref{sec:forced_hit_existing_methods} cannot be applied to this situation. In contrast, the present method is available because it relies solely on a simple assumption, namely, the sensor motion on the sea surface, $d \boldsymbol{x}_s /dt = \boldsymbol{u}_f ( \boldsymbol{x}_s )$. Thus, the comparison with AGW, which is introduced in \S~\ref{sec:forced_hit_existing_methods}, is conducted here, and the results show the present method outperforms AGW with a small number of sensors.

Additionally, in figure~\ref{fig:ocean_n_sensors_results}~(\textit{c}--\textit{d}), the major structure of the velocity fields (e.g., the eastward current around $35 ^\circ \textrm{N}$) can be roughly estimated with only eight sensors. It can be because ocean currents have steady flow structures and the present method can learn such structures through the training process, as mentioned in \S~\ref{sec:forced_hit_number_of_sensors}. This demonstration suggests that it may be possible to estimate ocean current fields by using a small number of buoys equipped with GPS sensors.

\subsubsection{Robustness to noise}
\label{sec:ocean_currents_robustness_to_noise}
\begin{figure}
  \centerline{
    \includegraphics[width=150mm]{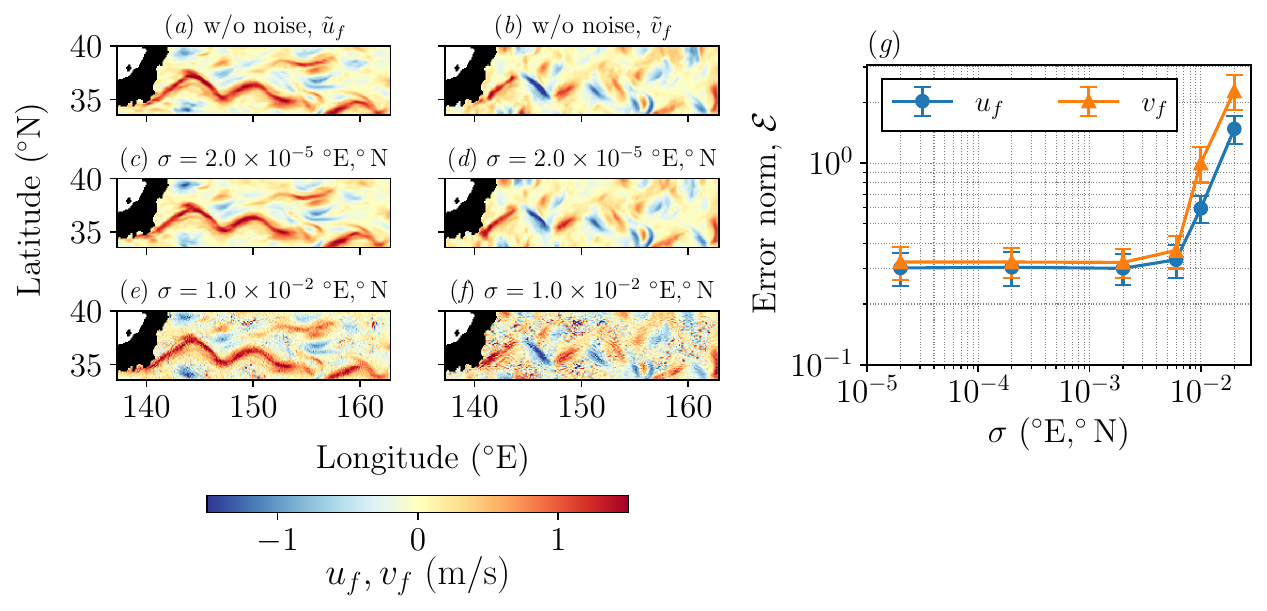}
  }
  \caption{(\textit{a}--\textit{f}) The estimated velocity fields of the ocean currents depending on the noise intensity $\sigma$. (\textit{g}) The error norm.}
  \label{fig:ocean_currents_noise}
\end{figure}
As a specific example, the noise robustness of sensors for the ocean currents is demonstrated in figure~\ref{fig:ocean_currents_noise}. The position uncertainty of GPS sensors is currently $\sim 2$~m~\cite{waas_global_2021}. To evaluate the noise influence, the Gaussian random noise, whose standard deviation is $\sigma$, is applied to the data of sensor locations. The noise intensity is set in the range of $\sigma = 2 \times 10^{-5}$\,--\,$2 \times 10^{-2} \ ^\circ \textrm{E}, ^\circ \textrm{N}$, where $1 ^\circ \textrm{E}, ^\circ \textrm{N} \simeq 1 \times 10^{5}$~m in the latitudes of $33.6$\,--\,$39.9 ^\circ \textrm{N}$. Figure~\ref{fig:ocean_currents_noise}~(\textit{c}--\textit{d}) shows that the noise with $\sigma = 2 \times 10^{-5} \ ^\circ \textrm{E}, ^\circ \textrm{N}$, which corresponds to $\sigma \simeq 2$~m, does not affect the estimation accuracy. The estimated fields become noisy when the noise intensity $\sigma$ is larger than $1 \times 10^{-2} \ ^\circ \textrm{E}, ^\circ \textrm{N} \simeq 1$~km, as shown in figure~\ref{fig:ocean_currents_noise}~(\textit{e}--\textit{f}) and (\textit{g}). This noise intensity corresponds to $\sigma = 0.1 \Delta x$, where $\Delta x$ is the grid size of the flow field, similarly to the case of the forced HIT discussed in \S~\ref{sec:forced_hit_robustness_to_noise}. In sum, the present method is robust enough in terms of the position uncertainty for the application of ocean current measurements using GPS sensors.

\section{Conclusions}
\label{sec:conclusions}

We have proposed a novel method to estimate flow velocity fields using only floating sensor locations based on machine learning (ML) techniques. The ML model is supposed to generate the estimated velocity fields so that the time variation of sensor locations in those fields is consistent with the given data. This method can be used without knowing either ground-truth velocity fields or governing equations for fluid flows, which is attractive for practical applications. The present method is validated using three examples: the two-dimensional flow around a circular cylinder, the two-dimensional forced homogeneous isotropic turbulence, and the two-dimensional ocean currents. The estimation performance and practicality are demonstrated through the investigations of the dependence on the number of sensors, time intervals, and the noise in the sensor location data. These results and the comparison with the existing methods show that the present method has excellent performance and practical robustness without any assumption for governing equations for fluid flows. Furthermore, the validation reveals that the present method can estimate the major flow structures with only a few sensors. We consider that it is because the ML model can learn the steady flow structures over the entire dataset of sensor locations. 

In future, we aim to apply the present method to the reconstruction of three-dimensional flow fields.We also believe that the present method would be further promised if the validation using actual floating sensors could be experimentally conducted.

\section*{Acknowledgments}
This work was supported through JSPS KAKENHI (Grant No. 21H05007) by Japan Society for the Promotion of Science.

\section*{Data availability statement}
The sample codes of this study are openly available on the authors' GitHub repository at \url{https://github.com/oura-tomoya/floating-sensors}.

\appendix

\section{Assessment of the model structure}\label{sec:appendix_model_structure}

\begin{table}
	\caption{The estimation accuracy dependence on the model structure.}
	\centering
	\begin{tabular}{ccccc}
	\toprule
	
	Activation function & \# of hidden layers & Kernel size & Error norm $\mathcal{E}$ of $u_f$ \\
	\hline
	ReLU & $4$ & $(7, 7)$ & $0.14 \pm 0.02$ \\
	Sigmoid & $4$ & $(7, 7)$ & $1.00 \pm 0.00$ \\
	tanh & $4$ & $(7, 7)$ & $0.17 \pm 0.02$ \\
	Softplus & $4$ & $(7, 7)$ & $0.14 \pm 0.02$ \\
	ReLU & $2$ & $(7, 7)$ & $0.15 \pm 0.02$ \\
	ReLU & $3$ & $(7, 7)$ & $0.16 \pm 0.02$ \\
	ReLU & $5$ & $(7, 7)$ & $0.14 \pm 0.02$ \\
	ReLU & $4$ & $(3, 3)$ & $0.22 \pm 0.03$ \\
	ReLU & $4$ & $(5, 5)$ & $0.17 \pm 0.03$ \\

	\bottomrule
	\end{tabular}
	\label{tab:model_structure}
\end{table}

To determine the ML model structure for the present method, the activation function, the number of hidden layers in the MLP part, and the kernel size of the CNN part are investigated. For this assessment, the two-dimensional HIT with $N_s = 512$ explained in \S \ref{sec:method_ocean_currents} is used. Table~\ref{tab:model_structure} shows the error norm $\mathcal{E}$ defined in equation (\ref{eq:normalized_error_norm}) dependence on the model structures. First, considering four activation functions, namely, the Rectified Linear Unit (ReLU), the sigmoid function, the hyperbolic tangent function, and the softplus function, ReLU and the softplus function show better accuracy whereas the sigmoid function completely fails to train the model and outputs uniformly zero-velocity fields. Second, changing the number of layers in the MLP does not result in any significant differences. Third, although a larger kernel size in the CNN part improves the estimation accuracy, the effect tends to be saturated. From these results, the model including ReLU as an activation function, four hidden layers in the MLP part, and the kernel size of $(7, 7)$ in the CNN part is adopted in this study.

\section{Divergence-free constraint}
\label{app:divergence_free_constraint}
\begin{figure}
  \centerline{
    \includegraphics[width=130mm]{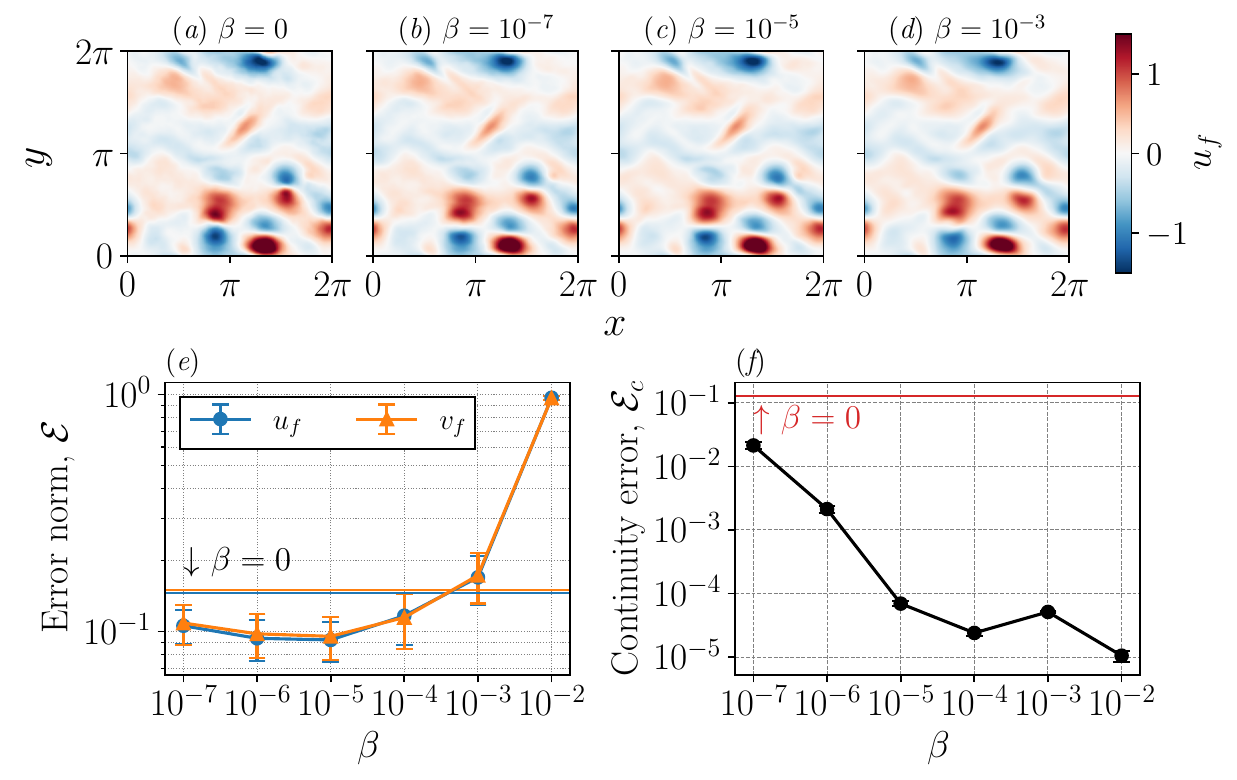}
  }
  \caption{(\textit{a}--\textit{d}) The estimated velocity fields of two-dimensional forced HIT with the divergence-free constraint depending on the constraint strength $\beta$. (\textit{e}) The error norm. The horizontal solid lines show the error norm without the constraint. (\textit{f}) The continuity error. The horizontal solid line shows the error without the constraint.}
  \label{fig:continuity_loss}
\end{figure}
Since the present method does not assume the governing equation for fluid fields, the estimated flow fields are not guaranteed to satisfy the divergence-free condition. We believe this is one of the properties of our method in that it can be used for non-divergence-free flows, but it can also be modified to take into account the continuity equation. Here, the weight optimisation for model training is conducted with
\begin{equation}
  \boldsymbol{w} = \textrm{argmin}_{\boldsymbol{w}} \left( \left\| \boldsymbol{x}_s^{n+1, i} - \tilde{\boldsymbol{x}}_s^{n+1, i} \right\|_2^2 + \beta \left\| \nabla \cdot \tilde{\boldsymbol{u}}_f^{n} \right\|_2^2 \right)
  ,
  \label{eq:continutiy_loss}
\end{equation}
where $\beta$ is the parameter to determine the strength of the constraint, instead of equation~(\ref{eq:weight_optimisation}).

Figure~\ref{fig:continuity_loss} shows the estimation results of the two-dimensional forced HIT depending on $\beta$. In figure~\ref{fig:continuity_loss} (\textit{f}), the continuity error is defined as
\begin{equation}
  \mathcal{E}_c = \left\| \nabla \cdot \tilde{\boldsymbol{u}}_f \right\|_2^2
  \label{eq:continutiy_error}
\end{equation}
for quantitative comparison. Because the constraint is implemented as the weak enforcement of the incompressible continuity equation, the estimation accuracy and the continuity error depend on the constraint strength $\beta$, and these are in a trade-off relationship. The best parameter to balance $\mathcal{E}$ and $\mathcal{E}_c$ appears to be around $\beta = 10^{-5}$ in this case. When the suitable parameter $\beta$ is utilized, both the estimation accuracy and the continuity error become better than those without the constraint. 

Although the divergence-free constraint yields favorable results as above, two points should be noted here. First, the assumption of the incompressible continuity equation does not hold in certain cases, such as compressible flow and flow with sources and sinks. Second, as shown in figure~\ref{fig:continuity_loss}, the parameter $\beta$ has to be determined as a hyper-parameter. For these reasons, we believe that the method without any governing equations for fluid flows, namely, without both the continuity equation and the Navier--Stokes equations is more versatile for practical applications.

\bibliographystyle{unsrtnat}
\bibliography{references}

\begin{thebibliography}{39}
\providecommand{\natexlab}[1]{#1}
\providecommand{\url}[1]{\texttt{#1}}
\expandafter\ifx\csname urlstyle\endcsname\relax
  \providecommand{\doi}[1]{doi: #1}\else
  \providecommand{\doi}{doi: \begingroup \urlstyle{rm}\Url}\fi

\bibitem[Wong et~al.(2020)]{wong_argo_2020}
A.~P.~S. Wong et~al.
\newblock Argo data 1999^^e2^^80^^932019: Two million temperature-salinity profiles and subsurface velocity observations from a global array of profiling floats.
\newblock \emph{Front. Mar. Sci.}, 7:\penalty0 700, 2020.

\bibitem[Hansen and Poulain(1996)]{hansen_quality_1996}
D.~V. Hansen and P-M. Poulain.
\newblock Quality control and interpolations of {WOCE}-{TOGA} drifter data.
\newblock \emph{J. Atmos. Ocean. Technol.}, 13:\penalty0 900--909, 1996.

\bibitem[Mokhasi et~al.(2009)Mokhasi, Rempfer, and Kandala]{mokhasi_predictive_2009}
P.~Mokhasi, D.~Rempfer, and S.~Kandala.
\newblock Predictive flow-field estimation.
\newblock \emph{Phys. D: Nonlinear Phenom.}, 238:\penalty0 290--308, 2009.

\bibitem[Podvin et~al.(2018)Podvin, Nguimatsia, Foucaut, Cuvier, and Fraigneau]{podvin_combining_2018}
B.~Podvin, S.~Nguimatsia, J-M. Foucaut, C.~Cuvier, and Y.~Fraigneau.
\newblock On combining linear stochastic estimation and proper orthogonal decomposition for flow reconstruction.
\newblock \emph{Exp. Fluids}, 59:\penalty0 58, 2018.

\bibitem[Wunsch and Heimbach(2007)]{wunsch_practical_2007}
C.~Wunsch and P.~Heimbach.
\newblock Practical global oceanic state estimation.
\newblock \emph{Phys. D: Nonlinear Phenom.}, 230:\penalty0 197--208, 2007.

\bibitem[Tossavainen et~al.(2008)Tossavainen, Percelay, Tinka, Wu, and Bayen]{tossavainen_ensemble_2008}
O-P. Tossavainen, J.~Percelay, A.~Tinka, A.~Wu, and A.~M. Bayen.
\newblock Ensemble {Kalman} filter based state estimation in {2D} shallow water equations using {Lagrangian} sensing and state augmentation.
\newblock In \emph{2008 47th IEEE Conference on Decision and Control}, pages 1783--1790, 2008.

\bibitem[Tinka et~al.(2013)Tinka, Rafiee, and Bayen]{tinka_floating_2013}
A.~Tinka, M.~Rafiee, and A.~M. Bayen.
\newblock Floating sensor networks for river studies.
\newblock \emph{IEEE Syst. J.}, 7:\penalty0 36--49, 2013.

\bibitem[Callaham et~al.(2019)Callaham, Maeda, and Brunton]{callaham_robust_2019}
J.~L. Callaham, K.~Maeda, and S.~L. Brunton.
\newblock Robust flow reconstruction from limited measurements via sparse representation.
\newblock \emph{Phys. Rev. Fluids}, 4:\penalty0 103907, 2019.

\bibitem[Erichson et~al.(2020)Erichson, Mathelin, Yao, Brunton, Mahoney, and Kutz]{erichson_shallow_2020}
N.~B. Erichson, L.~Mathelin, Z.~Yao, S.~L. Brunton, M.~W. Mahoney, and J.~N. Kutz.
\newblock Shallow neural networks for fluid flow reconstruction with limited sensors.
\newblock \emph{Proc. R. Soc. A}, 476:\penalty0 20200097, 2020.

\bibitem[Fukami et~al.(2021)Fukami, Maulik, Ramachandra, Fukagata, and Taira]{fukami_global_2021}
K.~Fukami, R.~Maulik, N.~Ramachandra, K.~Fukagata, and K.~Taira.
\newblock Global field reconstruction from sparse sensors with {Voronoi} tessellation-assisted deep learning.
\newblock \emph{Nat. Mach. Intell.}, 3:\penalty0 945--951, 2021.
\newblock ISSN 2522-5839.

\bibitem[{LeCun} et~al.(1998){LeCun}, Bottou, Bengio, and Haffner]{y_lecun_gradient-based_1998}
Y.~{LeCun}, L.~Bottou, Y.~Bengio, and P.~Haffner.
\newblock Gradient-based learning applied to document recognition.
\newblock \emph{Proc. IEEE}, 86:\penalty0 2278--2324, 1998.

\bibitem[Clark Di~Leoni et~al.(2023)Clark Di~Leoni, Agarwal, Zaki, Meneveau, and Katz]{clark_di_leoni_reconstructing_2023}
P.~Clark Di~Leoni, K~Agarwal, T.~A. Zaki, C.~Meneveau, and J.~Katz.
\newblock Reconstructing turbulent velocity and pressure fields from under-resolved noisy particle tracks using physics-informed neural networks.
\newblock \emph{Exp. Fluids}, 64:\penalty0 95, 2023.

\bibitem[Raissi et~al.(2019)Raissi, Perdikarism, and Karniadakis]{raissi_physics_2019}
M.~Raissi, P.~Perdikarism, and G.~E. Karniadakis.
\newblock Physics-informed neural networks: {A} deep learning framework for solving forward and inverse problems involving nonlinear partial differential equations.
\newblock \emph{J. Comput. Phys.}, 378:\penalty0 686--707, 2019.

\bibitem[Cai et~al.(2020)Cai, Liang, Gao, Xu, and Wei]{cai_particle_2020}
S.~Cai, J.~Liang, Q.~Gao, C.~Xu, and R.~Wei.
\newblock Particle image velocimetry based on a deep learning motion estimator.
\newblock \emph{IEEE Trans. Instrum. Meas.}, 69:\penalty0 3538--3554, 2020.

\bibitem[Morimoto et~al.(2021)Morimoto, Fukami, and Fukagata]{morimoto_experimental_2021}
M.~Morimoto, K.~Fukami, and K.~Fukagata.
\newblock Experimental velocity data estimation for imperfect particle images using machine learning.
\newblock \emph{Phys. Fluids}, 33:\penalty0 087121, 2021.

\bibitem[Zhang et~al.(2023{\natexlab{a}})Zhang, Nie, Dong, and Sun]{zhang_pyramidal_2023}
W.~Zhang, X.~Nie, X.~Dong, and Z.~Sun.
\newblock Pyramidal deep-learning network for dense velocity field reconstruction in particle image velocimetry.
\newblock \emph{Exp. Fluids}, 64:\penalty0 12, 2023{\natexlab{a}}.

\bibitem[Hasanuzzaman et~al.(2023)Hasanuzzaman, Eivazi, Merbold, Egbers, and Vinuesa]{Hasanuzzaman_enhancement_2023}
G.~Hasanuzzaman, H.~Eivazi, S.~Merbold, C.~Egbers, and R.~Vinuesa.
\newblock Enhancement of {PIV} measurements via physics-informed neural networks.
\newblock \emph{Meas. Sci. Technol.}, 34:\penalty0 044002, jan 2023.

\bibitem[Zhang et~al.(2023{\natexlab{b}})Zhang, Dong, Sun, and Xu]{zhang_unsupervised_2023}
W.~Zhang, X.~Dong, Z.~Sun, and S.~Xu.
\newblock An unsupervised deep learning model for dense velocity field reconstruction in particle image velocimetry ({PIV}) measurements.
\newblock \emph{Phys. Fluids}, 35:\penalty0 077108, 2023{\natexlab{b}}.

\bibitem[Rumelhart et~al.(1986)Rumelhart, Hinton, and Williams]{rumelhart1986learning}
D.~E. Rumelhart, G.~E. Hinton, and R.~J. Williams.
\newblock Learning representations by back-propagating errors.
\newblock \emph{Nature}, 323:\penalty0 533--536, 1986.

\bibitem[Kingma and Ba(2017)]{kingma2017adam}
D.~P. Kingma and J.~Ba.
\newblock Adam: A method for stochastic optimization, 2017.
\newblock (arXiv:1412.6980).

\bibitem[Kor et~al.(2017)Kor, Badri~Ghomizad, and Fukagata]{kor_unified_2017}
H~Kor, M.~Badri~Ghomizad, and K.~Fukagata.
\newblock A unified interpolation stencil for ghost-cell immersed boundary method for flow around complex geometries.
\newblock \emph{J. Fluid Sci. Technol.}, 12:\penalty0 JFST0011, 2017.

\bibitem[Charney(1971)]{charney_geostrophic_1971}
J.~G. Charney.
\newblock Geostrophic turbulence.
\newblock \emph{J. Atmos. Sci.}, 28:\penalty0 1087--1095, 1971.

\bibitem[Xia et~al.(2009)Xia, Shats, and Falkovich]{xia_spectrally_2009}
H.~Xia, M.~Shats, and G.~Falkovich.
\newblock Spectrally condensed turbulence in thin layers.
\newblock \emph{Phys. Fluids}, 21:\penalty0 125101, 2009.

\bibitem[San and Staples(2013)]{san_stationary_2013}
O.~San and A.~E. Staples.
\newblock Stationary two-dimensional turbulence statistics using a {Markovian} forcing scheme.
\newblock \emph{Comput. Fluids}, 71:\penalty0 1--18, 2013.

\bibitem[Lundgren(2003)]{lundgren_linearly_2003}
T.~S. Lundgren.
\newblock Linearly forced isotropic turbulence.
\newblock Technical report, Center for Turbulence Research, Stanford, 2003.

\bibitem[Maltrud and Vallis(1991)]{maltrud_energy_1991}
M.~E. Maltrud and G.~K. Vallis.
\newblock Energy spectra and coherent structures in forced two-dimensional and beta-plane turbulence.
\newblock \emph{J. Fluid Mech.}, 228:\penalty0 321--342, 1991.

\bibitem[Schorghofer(2000)]{schorghofer_energy_2000}
N~Schorghofer.
\newblock Energy spectra of steady two-dimensional turbulent flows.
\newblock \emph{Phys. Rev. E}, 61:\penalty0 6572--6577, 2000.

\bibitem[Taira et~al.(2016)Taira, Nair, and Brunton]{taira_network_2016}
K.~Taira, A.~G. Nair, and S.~L. Brunton.
\newblock Network structure of two-dimensional decaying isotropic turbulence.
\newblock \emph{J. Fluid Mech.}, 795:\penalty0 R2, 2016.

\bibitem[Masumoto et~al.(2004)]{masumoto_fifty_2004}
Y.~Masumoto et~al.
\newblock A fifty-year eddy-resolving simulation of the world ocean: Preliminary outcomes of {OFES} ({OGCM} for the earth simulator).
\newblock \emph{J. Earth Simulator}, 1:\penalty0 35--56, 2004.

\bibitem[Sasaki et~al.(2004{\natexlab{a}})Sasaki, Sasai, Kawahara, Furuichi, Araki, Ishida, Yamanaka, Masumoto, and Sakuma]{sasaki_series_2004}
H.~Sasaki, Y.~Sasai, S.~Kawahara, M.~Furuichi, F.~Araki, A.~Ishida, Y.~Yamanaka, Y.~Masumoto, and H.~Sakuma.
\newblock A series of eddy-resolving ocean simulations in the world ocean - {OFES} ({OGCM} for the earth simulator) project.
\newblock In \emph{OCEANS `04}, pages 1535--1541, 2004{\natexlab{a}}.

\bibitem[Sasaki et~al.(2004{\natexlab{b}})Sasaki, Ishida, Yamanaka, and Sasaki]{sasaki_chloro_2004}
Y.~Sasaki, A.~Ishida, Y.~Yamanaka, and H.~Sasaki.
\newblock Chlorofluorocarbons in a global ocean eddy-resolving {OGCM}: Pathway and formation of antarctic bottom water.
\newblock \emph{Geophysical Research Letters}, 31, 2004{\natexlab{b}}.

\bibitem[Sasaki et~al.(2008)Sasaki, Nonaka, Masumoto, Sasai, Uehara, and Sakuma]{sasaki_eddy_2008}
H.~Sasaki, M.~Nonaka, Y.~Masumoto, Y.~Sasai, H.~Uehara, and H.~Sakuma.
\newblock \emph{An Eddy-Resolving Hindcast Simulation of the Quasiglobal Ocean from 1950 to 2003 on the Earth Simulator}, pages 157--185.
\newblock Springer New York, 2008.

\bibitem[Leith(1968)]{leith_diffusion_1968}
C.~E. Leith.
\newblock Diffusion approximation for two-dimensional turbulence.
\newblock \emph{Phys. Fluids}, 11:\penalty0 671--672, 1968.

\bibitem[Legras et~al.(1988)Legras, Santangelo, and Benzi]{legras_high-resolution_1988}
B.~Legras, P.~Santangelo, and R.~Benzi.
\newblock High-resolution numerical experiments for forced two-dimensional turbulence.
\newblock \emph{Europhys. Lett.}, 5:\penalty0 37--42, 1988.

\bibitem[Pathak et~al.(2016)Pathak, Kr\"ahenb\"uhl, Donahue, Darrell, and Efros]{pathak_context_2016}
D.~Pathak, P.~Kr\"ahenb\"uhl, J.~Donahue, T.~Darrell, and A.~A. Efros.
\newblock Context encoders: Feature learning by inpainting.
\newblock In \emph{Computer Vision and Pattern Recognition ({CVPR})}, 2016.

\bibitem[Iizuka et~al.(2017)Iizuka, Simo-Serra, and Ishikawa]{iizuka_globally_2017}
S.~Iizuka, E.~Simo-Serra, and H.~Ishikawa.
\newblock Globally and locally consistent image completion.
\newblock \emph{ACM Trans. Graph.}, 36:\penalty0 1--14, 2017.

\bibitem[Ag\"{u}\'{i} and Jim\'{e}nez(1987)]{agui_performance_1987}
J.~C. Ag\"{u}\'{i} and J.~Jim\'{e}nez.
\newblock On the performance of particle tracking.
\newblock \emph{J. Fluid Mech.}, 185:\penalty0 447--468, 1987.

\bibitem[Clark Di~Leoni(2022)]{clark_pinns_2022}
P.~Clark Di~Leoni.
\newblock {PINNs}.
\newblock 2022.
\newblock URL \url{{https://github.com/PatricioClark/PINNs}}.

\bibitem[{WAAS T\&E Team}(2021)]{waas_global_2021}
{WAAS T\&E Team}.
\newblock Global positioning system standard positioning service performance analysis report.
\newblock Technical Report 112, {FAA William J. Hughes Technical Center}, 2021.

\end{thebibliography}

\end{document}